\documentclass[preprint,amsmath,amssymb,aps,nofootinbib,showpacs]{revtex4}
\usepackage{graphicx}
\usepackage{bm}
\usepackage{subfigure}
\usepackage{amsmath}

\newcommand{\epl}{Europhys. Lett.\ }

\newcommand{\pr}{Phys. Rev.\ }

\newcommand{\jpa}{J. Phys. A\ }
\newcommand{\jpb}{J. Phys. B\ }
\newcommand{\njp}{New J. Phys.\ }
\newcommand{\etal}{{\em et al. }}
\newcommand{\e}{\mbox{e}}
\newcommand{\tr}{\mbox{tr}}

\newcommand{\UQ}{School of Mathematics and Physics, University of Queensland, Brisbane, 
QLD 4072, Australia.}

\newcommand{\UFF}{Instituto da F\'{\i}sica da Universidade 
Federal Fluminense, Boa Viagem 24210-340, 
Niter\'oi - RJ, Brazil}

\begin{document}

\title{Two-well atomic Bose-Hubbard analogues of optical cavities}

\author{M.~K. Olsen and C.~V. Chianca}
\affiliation{\UQ}

\author{K. Dechoum}
\affiliation{\UFF}

\date{\today}

\begin{abstract}

We propose and analyse analogs of optical cavities for atoms using two-well Bose-Hubbard models with pumping and losses.
With one well pumped, we find that both the mean-field dynamics and the quantum statistics show a quantitative dependence on the choice of damped well. Both the systems we analyse remain far from equilibrium, preserving good coherence between the wells in the steady-state. We find a small degree of quadrature squeezing and mode entanglement for some parameter regimes. Due to recent experimental advances, it should be possible to demonstrate the effects we investigate and predict.

\end{abstract}


\maketitle

\section{Introduction}
\label{sec:intro}

Recent advances in the techniques of configuring optical potentials~\cite{painting,tylerpaint} allow for the fabrication of lattice potentials for ultra-cold atoms in a variety of geometric configurations. Combined with the technique of causing dissipation from a particular well through use of an electron beam~\cite{NDC} or by optical means~\cite{Weitenberg}, and the possibility of pumping a Bose-Hubbard system from a larger reservoir condensate~\cite{Kordas1,Kordas2}, we have the elements required for the fabrication of nonlinear damped and pumped optical cavities with varying configurations. In this work we perform theoretical investigations of two different Bose-Hubbard models~\cite{BHmodel,Jaksch,BHJoel} with added pumping and loss. We investigate the population dynamics, the quantum statistics of the system such as squeezing and inseparability, and a pseudo-entropy obtained from a reduced single particle density matrix. We will show that there is some degree of steady-state quadrature squeezing in both the configurations we examine, that mode inseparability is demonstrated using quadrature measures, and that the population dynamics and quantum statistical fatures depend on both the configuration and the collisional nonlinearity.  

An early investigation by Drummond and Walls analysed a quantum optical system consisting of a Kerr medium inside a Fabry-Perot cavity, which is mathematically the equivalent of a pumped and damped single well Bose-Hubbard model~\cite{PDDDFW}, with the main difference being that Kerr nonlinearities tend to be higher with atomic systems. More recently, Pi\u{z}orn has analysed Bose-Hubbard models with pumping and dissipation~\cite{Pizorn}, using density matrix techniques, which are useful for moderate numbers of atoms and wells. Boit\'e \etal have analysed a two dimensional Bose-Hubbard model in terms of steady-state phases and instabilities, with an emphasis on coupled photonic microcavities~\cite{Boite}. More recently, Cui \etal have investigated driven and dissipative Bose-Hubbard models, obtaining mean-field analytical results for a two-well system~\cite{Cui}.  In this work we analyse both the dynamics and steady-state properties of our systems, going beyond the mean-field approximation with the truncated Wigner representation~\cite{Graham,Steel}, which does not impose a computational limitation on the number of atoms. The main advantages of the truncated Wigner representation are that the computational complexity scales linearly with the number of wells, and it does not suffer from the catastrophic instabilities of the positive-P representation~\cite{P+}.

\section{Physical model, Hamiltonian, and equations of motion}
\label{sec:model}

In this investigation we use the truncated Wigner representation~\cite{Graham,Steel}, which we fully expect to be accurate for our systems in the presence of pumping and dissipation. Although this method will not capture any revivals in population oscillations in an isolated Bose-Hubbard dimer~\cite{Chiancathermal}, nor will it calculate two-time correlation functions accurately~\cite{turco2time}, we do not expect the first in a damped system, and we are not interested in the second here. The truncated Wigner representation goes beyond the pairing mean-field theory~\cite{PMFT} and the Bogoliubov back reaction method~\cite{BBRVardi1,BBRVardi2} previously used in theoretical analyses in that it imposes no factorisation assumptions on correlations, irrespective of their order. 

Beginning with the two-well unitary Bose-Hubbard Hamiltonian, this is written as
\begin{equation}
{\cal H} = \hbar\chi\sum_{i=1}^{2}\hat{a}_{i}^{\dag\,2}\hat{a}_{i}^{2}-\hbar J \left(\hat{a}_{1}^{\dag}\hat{a}_{2}+\hat{a}_{2}^{\dag}\hat{a}_{1} \right),
\label{eq:genHam}
\end{equation}
where $\hat{a}_{i}$ is the bosonic annihilation operator for the $i$th well, $\chi$ represents the collisional nonlinearity and $J$ is the tunneling strength. We will always consider that the pumping is into well $1$, which can be represented by the Hamiltonian
\begin{equation}
{\cal H}_{pump} = i\hbar\left(\hat{\Gamma}\hat{a}_{1}^{\dag}-\hat{\Gamma}^{\dag}\hat{a}_{1}\right),
\label{eq:pump}
\end{equation}
which is commonly used for the investigation of optical cavities. The basic assumption here is that the first well receives atoms from a coherent condensate which is much larger than any of the modes in the wells we are investigating, so that it will not become noticeably depleted over the time scales of interest.
The damping term for well $i$ acts on the system density matrix as the Lindblad superoperator
\begin{equation}
{\cal L}\rho = \gamma\left(2\hat{a}_{i}\rho\hat{a}_{i}^{\dag}-\hat{a}_{i}^{\dag}\hat{a}_{i}\rho-\rho\hat{a}_{i}^{\dag}\hat{a}_{i}\right),
\label{eq:damp}
\end{equation}
where $\gamma$ is the coupling between the damped well and the atomic bath, which we assume to be unpopulated. Physically, such a damping process can be realised using an electron beam~\cite{NDC}.
If the lost atoms fall under gravity, we are justified in using the Markov and Born approximations~\cite{JHMarkov}.

Following the usual procedures~\cite{QNoise,DFW}, we may map the problem onto a generalised Fokker-Planck equation (FPE) for the Wigner distribution of the system. Since this generalised FPE contains third-order derivatives, we truncate at second order. Although it is possible to map the third-order derivatives onto stochastic difference equations, these are highly unstable~\cite{nossoEPL}. Having discarded these derivatives, we may map the resulting FPE onto It\^o stochastic equations~\cite{SMCrispin} for the Wigner variables. These equations for a two-well chain with pumping at well $1$ and loss at well $2$ are
\begin{eqnarray}
\frac{d\alpha_{1}}{dt} &=& \epsilon - 2i\chi |\alpha_{1}|^{2}\alpha_{1}+iJ\alpha_{2}. \nonumber \\
\frac{d\alpha_{2}}{dt} &=& -\gamma\alpha_{2}-2i\chi|\alpha_{2}|^{2}\alpha_{2}+iJ\alpha_{1}+\sqrt{\gamma}\xi,
\label{eq:BHp1g2}
 \end{eqnarray}
 with those with loss at the pumped well resulting from moving the terms proportional to $\gamma$. In the above equation, $\epsilon$ represents the rate at which atoms enter well $1$ from the pumping mode, $\gamma$ is the loss rate from the second well, and $\xi$ is a complex Gaussian noise with the moments $\overline{\xi(t)}=0$ and $\overline{\xi^{\ast}(t)\xi(t')}=\delta(t-t')$, where the upper line represents a classical averaging process. The variables $\alpha_{i}$ correspond to the operators $\hat{a}_{i}$ in the sense that averages of products of the Wigner variables over many stochastic trajectories become equivalent to symmetrically ordered operator expectation values, for example $\overline{|\alpha_{i}|^{2}}=\frac{1}{2}\langle\hat{a}_{i}^{\dag}\hat{a}_{i}+\hat{a}_{i}\hat{a}_{i}^{\dag}\rangle$. The initial states in all wells will be vacuum, sampled as in Olsen and Bradley~\cite{states} for coherent states with vacuum excitation. We note here that we will use $\epsilon=10$ and $\gamma=J=1$ in all our numerical investigations, while varying the value of $\chi$. We have averaged over at least $3\times 10^{5}$ stochastic trajectories for all the graphical results presented here, and the sampling error is typically less than the plotted line widths. 
  
\section{Quantities of interest}
\label{sec:interest}

There are several quantities worthy of investigation here, including the populations in each well, $\overline{|\alpha_{i}|^{2}}-\frac{1}{2}$, the coherences between the wells, the currents into each well, the quadrature variances, a reduced single-particle pseudo-entropy, and measures of separability and entanglement. We firstly define the real coherence function between wells $1$ and $2$,
\begin{equation}
\sigma_{12} =\sqrt{ \langle \hat{a}_{1}^{\dag}\hat{a}_{2}\rangle\langle \hat{a}_{1}\hat{a}_{2}^{\dag}\rangle} .
\label{eq:sigmaij}
\end{equation}
Note that we define this as a real function so that it may be plotted, which is not as simple for the actual complex coherence, and take the square root so that it will be of the same magnitude as the currents. If our atomic cavities behave as a collection of superfluid states analogous to the electromagnetic field in a pumped optical cavity without internal nonlinearity, we expect that these would obtain their coherent state values in the steady state, for example,
\begin{equation}
\sigma_{12}\rightarrow \overline{|\alpha_{1}||\alpha_{2}|}.
\label{sec:coherentsigma}
\end{equation}
The inclusion of finite $\chi$, with the attendant phase-diffusion~\cite{Lewenstein,Steel,Chiancathermal} and shearing of the Wigner function~\cite{nonGauss}, should act to decrease these values.
The current from well $1$ into well $2$ is defined as 
\begin{equation}
I_{12}=-i\langle\hat{a}_{2}^{\dag}\hat{a}_{1}-\hat{a}_{1}^{\dag}\hat{a}_{2}\rangle .
\label{eq:currentdef}
\end{equation}

Defining the atomic quadratures as 
\begin{eqnarray}
\hat{X}_{j}(\theta) &= & \hat{a}_{j}\e^{-i\theta}+\hat{a}_{j}^{\dag}\e^{i\theta},
\label{eqn:Xtheta}
\end{eqnarray}
so that $\hat{Y}_{j}(\theta)=\hat{X}_{j}(\theta+\pi/2)$, squeezing exists whenever a quadrature variance is found to be less than $1$, for any angle. As is well known, one of the effects of a $\chi^{(3)}$ nonlinearity is to cause any squeezing to be found at a non-zero quadrature angle~\cite{nlc}. Having defined our quadratures, we may now define the correlations we will investigate to detect entanglement between modes. The first of these, known as the Duan-Simon inequality~\cite{Duan,Simon}, states that, for any two separable states,
\begin{equation}
V(\hat{X}_{j}+\hat{X}_{k})+V(\hat{Y}_{j}-\hat{Y}_{k}) \geq 4,
\label{eq:DS}
\end{equation}
with any violation of this inequality demonstrating the inseparability of modes $j$ and $k$. 

A further set of inequalities, based on the Cauchy-Schwarz inequality, have been developed by Hillery and Zubairy~\cite{HZ}. They showed that, considering two separable modes denoted by $i$ and $j$,
\begin{equation}
| \langle \hat{a}_{i}^{\dag}\hat{a}_{j}\rangle |^{2} \leq \langle \hat{a}_{i}^{\dag}\hat{a}_{i}\hat{a}_{j}^{\dag}\hat{a}_{j}\rangle,
\label{eq:HZ}
\end{equation}
with the equality holding for coherent states. The violation of this inequality is thus an indication of the inseparability of, and entanglement between, the two modes. As shown by Olsen~\cite{BECsplit,BHspread,splitOC}, this is useful for systems where number conservation holds, and in which case the Duan-Simon criterion may not detect inseparability. Although this is not the case here, it is still of interest to compare the predictions with the quadrature inequalities defined above.
Using the Hillery-Zubairy result, we now define the correlation function
\begin{equation}
\xi_{12} = \langle \hat{a}_{1}^{\dag}\hat{a}_{2}\rangle\langle \hat{a}_{1}\hat{a}_{2}^{\dag}\rangle - \langle \hat{a}_{1}^{\dag}\hat{a}_{1}\hat{a}_{2}^{\dag}\hat{a}_{2}\rangle,
\label{eq:xiij}
\end{equation}
for which a positive value reveals entanglement between modes $1$ and $2$.  In the Wigner representation, this is found as
\begin{equation}
\xi_{12} = \overline{\alpha_{1}^{\ast}\alpha_{2}}\times\overline{\alpha_{2}^{\ast}\alpha_{1}}-\overline{|\alpha_{1}|^{2}|\alpha_{2}|^{2}}+\frac{1}{2}\left(\overline{|\alpha_{1}|^{2}+|\alpha_{2}|^{2}}\right)-\frac{1}{4}.
\label{eq:Wxi12}
\end{equation}

The last quantity which we investigate is a pseudo-entropy, derived from the single particle reduced density matrix~\cite{Anglin,BH4,Chiancathermal}, defined for two wells as 
\begin{equation}
{\cal R} = \frac{1}{\langle \hat{a}_{1}^{\dag}\hat{a}_{1}\rangle+\langle  \hat{a}_{2}^{\dag}\hat{a}_{2}\rangle}
\begin{bmatrix}
 \langle\hat{a}_{1}^{\dag}\hat{a}_{1}\rangle & \langle\hat{a}_{1}^{\dag}\hat{a}_{2}\rangle \\
\langle\hat{a}_{2}^{\dag}\hat{a}_{1}\rangle & \langle\hat{a}_{2}^{\dag}\hat{a}_{2}\rangle
 \end{bmatrix}.
\label{eq:Rmat}
\end{equation}
The pseudo-entropy is then defined in the standard von Neumann manner 
as
\begin{equation}
{\cal S} = -\tr\left({\cal R}\log{\cal R}\right).
\label{eq:vNS}
\end{equation}
Analytical values can be calculated in some limiting cases, such as a system of Fock or coherent states. These limiting cases are useful for the calculation of maximum values to which the system should relax if all coherences disappear. As a final note, we mention that all the quantities needed for the correlations above can in principle be measured, either by density (number) measurements or via atomic homodyning~\cite{andyhomo}. 

\section{Pumping and loss at different wells}
\label{sec:BH2b1g2}

For a two-well dimer, there are two different configurations that we investigate. The first has pumping at well $1$ with loss at well $2$, while the second has both pumping and loss at well $1$. As we show below, they exhibit qualitatively different behaviours.
The first configuration is described by Eq.~\ref{eq:BHp1g2}. If we set $\chi=0$, we can find the classical steady-state solutions,
\begin{eqnarray}
\alpha_{1} &=& \frac{\gamma\epsilon}{J^{2}}, \nonumber \\
\alpha_{2} &=& \frac{i\epsilon}{J},
\label{eq:classicp1g2}
\end{eqnarray}
showing that the steady-state populations of the two wells are equal for our parameters, with the coherent excitation of the first well being real while that of the second well is in the $\hat{Y}$ quadrature.

\begin{figure}[tbhp]
\includegraphics[width=0.75\columnwidth]{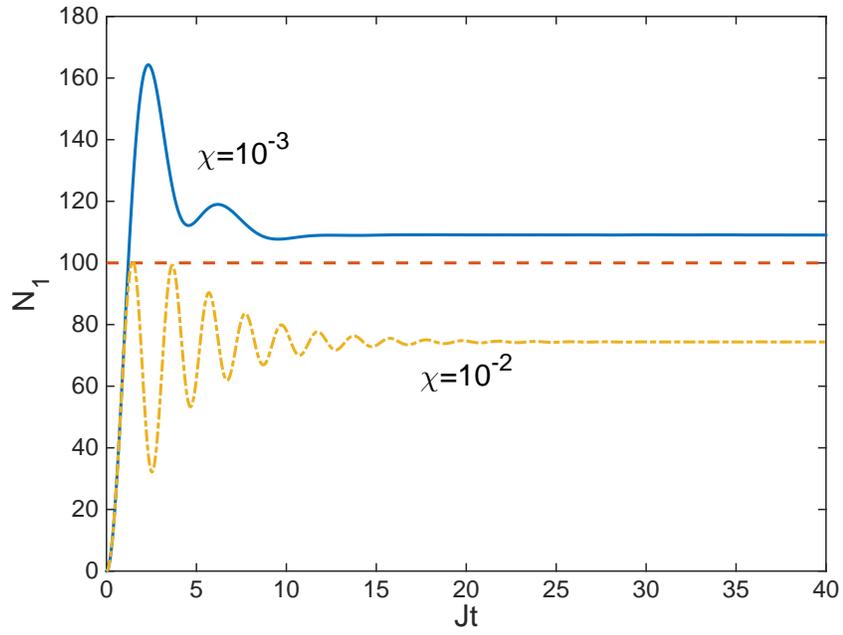}
\caption{(colour online) The populations of the first well for different $\chi$ values and loss at well $2$. The classical non-interacting value is shown by the dashed line. $Jt$ is a dimensionless time and all quantities plotted in this and subsequent plots are dimensionless.}
\label{fig:cav2g2N1}
\end{figure}

\begin{figure}[tbp]
\includegraphics[width=0.75\columnwidth]{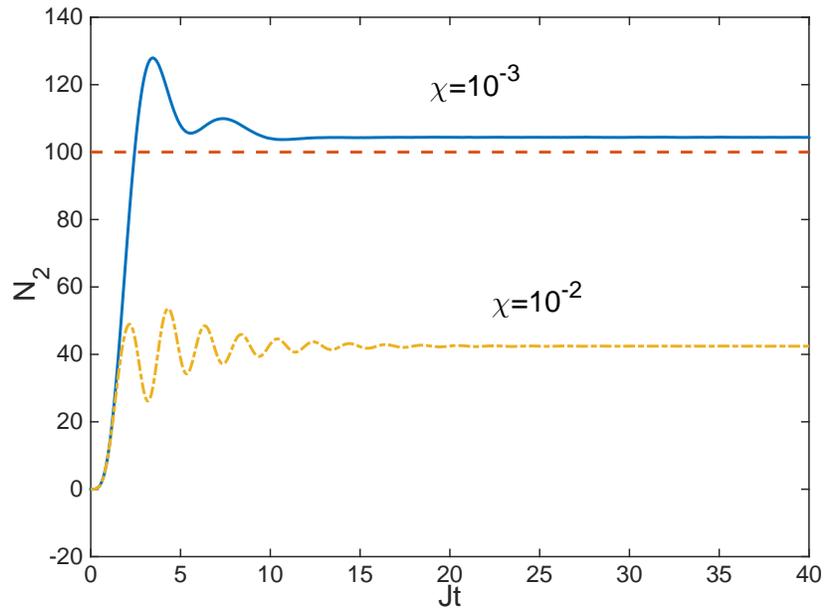}
\caption{(colour online) The populations of the second well for different $\chi$ values and loss at well $2$. The classical non-interacting value is shown by the dashed line.}
\label{fig:cav2g2N2}
\end{figure}

In Fig.~\ref{fig:cav2g2N1} we show the stochastically calculated populations in the first well, for $\chi=10^{-3}$ and $10^{-2}$. The classical non-interacting steady-state solution is shown as a dashed line. We see that, while the smaller value of $\chi$ causes the steady-state value to increase, the larger value causes it to decrease. The values for the second well are shown in Fig.~\ref{fig:cav2g2N2}, where we see the same trend, so that the total number of atoms decreases for the greater value of the nonlinearity. This is to be expected since the nonlinearity causes an imaginary component of the field analogous to that caused by detuning of an optical cavity, where the circulating power in an optical system decreases by a factor of $\gamma^{2}/(\gamma^{2}+\Delta^{2})$, where $\Delta$ is the detuning. The increase for the smaller $\chi$ value is counterintuitive and cannot be explained by the same reasoning.

\begin{figure}[tbhp]
\includegraphics[width=0.75\columnwidth]{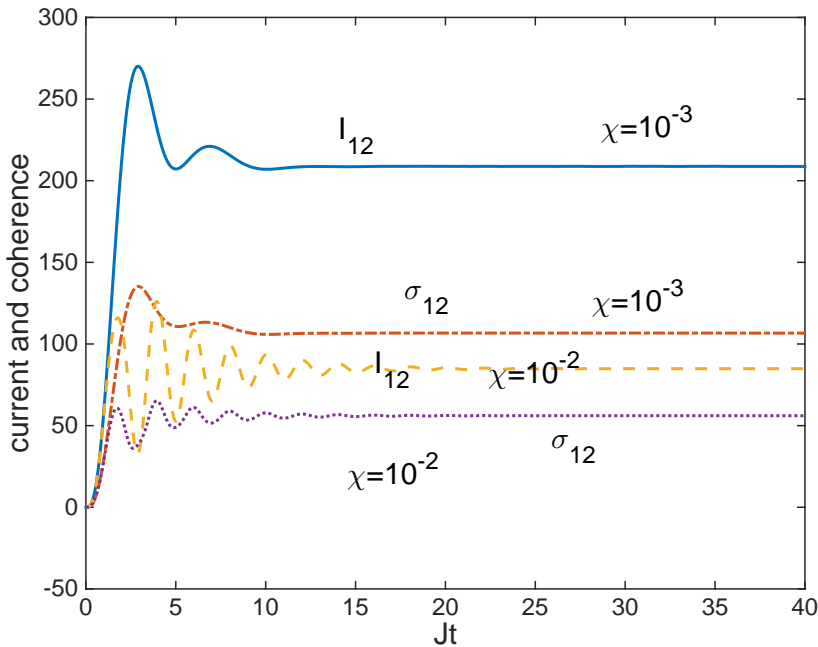}
\caption{(colour online) The currents from well $1$ into well $2$ and the coherence function, $\sigma_{12}$, for the two different values of $\chi$.}
\label{fig:cav2g2Isig}
\end{figure}

The currents into the second well and the coherence functions $\sigma_{12}$ are shown in Fig.~\ref{fig:cav2g2Isig}. We see that increasing the collisional nonlinearity decreases both the current and the coherences. A decrease in current can be explained by the fact that, with the higher nonlinearity, we are approaching the macroscopic self-trapping (MST) regime~\cite{BHJoel,Nemoto,Franzosi,Hines,Albiez}, where tunnelling is suppressed. The lower values of the coherences are explained almost entirely by the reduced populations, with phase diffusion playing a very limited role. Their values are almost indistinguishable from what is expected for two coherent states.

When we investigate the quantum statistics of the modes, we find steady-state quadrature squeezing and smallish violations of the Duan-Simon inequality of Eq.~\ref{eq:DS}. We present these values and the quadrature angles of the greatest violation in the table below.  
 We found that $\xi_{12}>0$ only in the transient regimes, with no steady-state violations of the Hillery-Zubairy inequality.

\begin{center}
 \begin{tabular}{||c || c  c || c  c ||} 
 \hline
  & $\chi=$ & $10^{-3}$ & $\chi=$ &  $10^{-2}$  \\ [0.5ex] 
 \hline\hline
 $V(\hat{X}_{1})$ & 0.65, & 20$^{o}$ & 0.62, & 122$^{o}$ \\ 
 \hline
 $V(\hat{X}_{2})$ & 0.78, & 102$^{o}$ & 0.69, & 2$^{o}$ \\ 
 \hline
 DS & 4.2, & 33$^{o}$ & 3.9, & 153$^{o}$ \\  
 \hline
\end{tabular}
\end{center}

The steady-state pseudo-entropy, ${\cal S}$, was found to be $0.02$ for $\chi=10^{-3}$ and $0.07$ for $\chi=10^{-2}$. These low values are a result of the persistence of the off-diagonal coherences in the steady-state, as can be seen in Eq.~\ref{eq:Rg2k3} and Eq.~\ref{eq:Rg2k2}. If the populations were equally distributed with no coherence between wells, we would find a value of $\log 2 \approx 0.6931$. For the actual mean populations, the values would be $0.6929$ for $\chi=10^{-3}$ and $0.6534$ for $\chi=10^{-2}$ if the coherences had disappeared. This is an indication that the populations in each well are close to coherent states.
The actual steady-state reduced density matrices are found as 
\begin{equation}
{\cal R}_{\chi=10^{-3}} = 
\begin{bmatrix}
0.51 & -0.10-0.49i \\
-0.10+0.49i & 0.49
 \end{bmatrix},
\label{eq:Rg2k3}
\end{equation}
and
\begin{equation}
{\cal R}_{\chi=10^{-2}} = 
\begin{bmatrix}
 0.64 & -0.31-0.36i \\
-0.31+0.36i & 0.36
 \end{bmatrix}.
\label{eq:Rg2k2}
\end{equation}

 \section{Pumping and loss at the same well}
\label{sec:BH2b1g1}

\begin{figure}[tbhp]
\includegraphics[width=0.75\columnwidth]{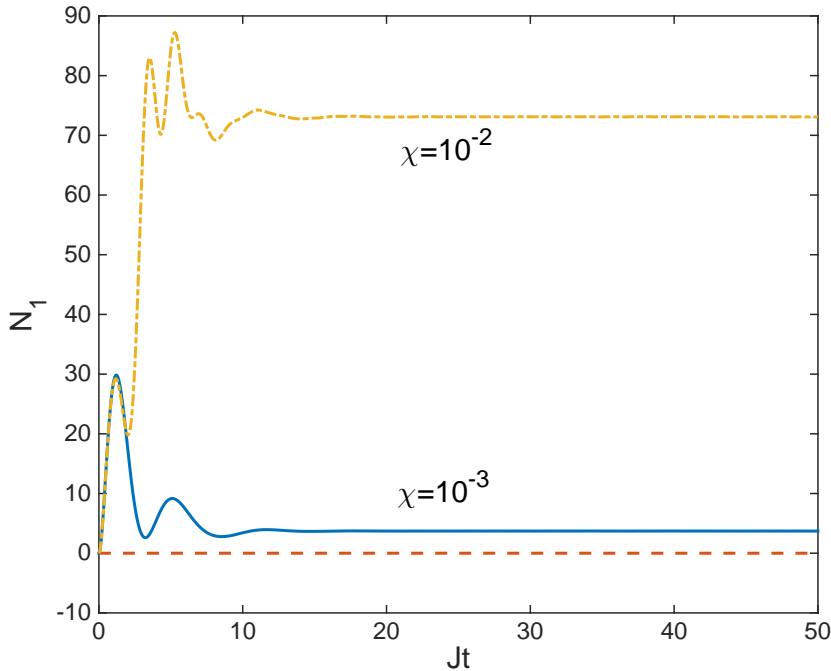}
\caption{(colour online) The populations of the first well for the two different different $\chi$ values and loss and pumping at well $1$. The dashed line represents the classical noninteracting prediction.}
\label{fig:BH2g1N1}
\end{figure}

\begin{figure}[tbhp]
\includegraphics[width=0.75\columnwidth]{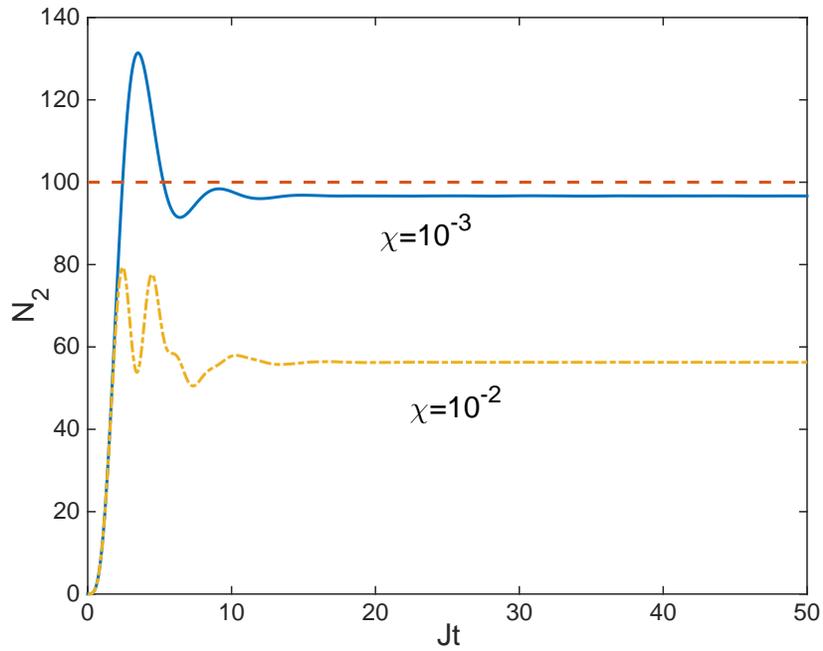}
\caption{(colour online) The populations of the second well for the two different $\chi$ values and loss and pumping at well $1$. The dashed line represents the classical noninteracting prediction.}
\label{fig:BH2g1N2}
\end{figure}

This configuration has both pumping and dissipation at the first well. The classical steady-state solutions with $\chi=0$ are found as
\begin{eqnarray}
\alpha_{1} &=& 0, \nonumber \\
\alpha_{2} &=& \frac{i\epsilon}{J},
\label{eq:classicp1g1}
\end{eqnarray}
so that the coherent excitation in the second well is again aligned with the $\hat{Y}$ quadrature. The first well, being vacuum, has no preferred phase. It is interesting that the first well remains unoccupied in the steady-state, with the tunneling between the two wells dropping to zero.
As can be seen from Fig.~\ref{fig:BH2g1N1}, the addition of a finite $\chi$ changes this so that well $1$ now has a non-zero steady-state occupation.
The population of the second well is decreased over the noninteracting value, for both values of $\chi$, as seen in Fig.~\ref{fig:BH2g1N2}.  We also see that the total steady-state mean occupation of the system is unchanged by $\chi=10^{-3}$, remaining at $100$ atoms. With $\chi=10^{-2}$, it increases to $129$, which is again counterintuitive. 

\begin{figure}[tbhp]
\includegraphics[width=0.75\columnwidth]{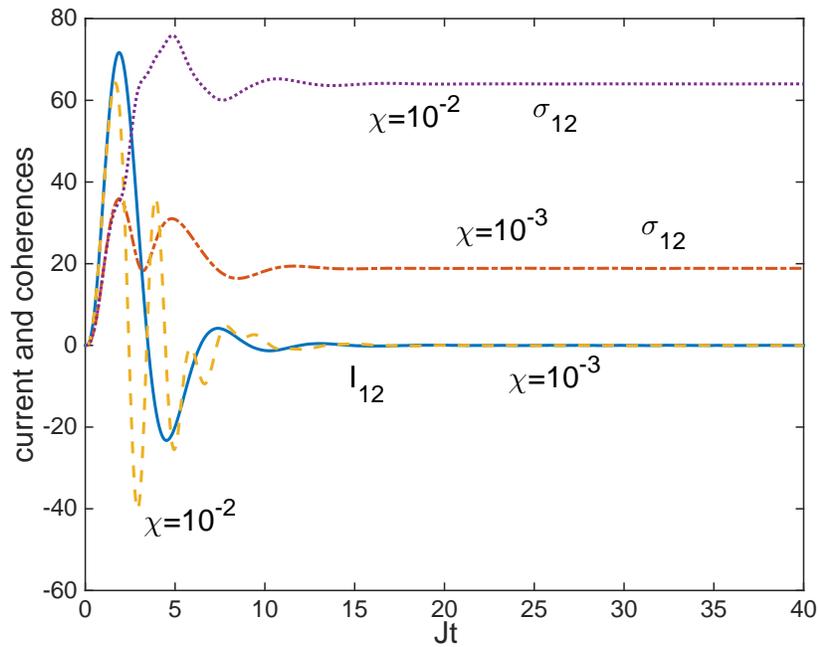}
\caption{(colour online) $I_{12}$ and $\sigma_{12}$ for the two different $\chi$ values, with both loss and pumping at well $1$.}
\label{fig:BH2g1Isig}
\end{figure}

In Fig.~\ref{fig:BH2g1Isig} we show the real coherence functions and the tunneling for this configuration. We see that the steady-state tunnelling goes to zero, which it must do to reach a state where the number in well $2$ remains constant. Once again the steady-state coherence functions are indistinguishable from their coherent state values.
The fact that the higher value of $\chi$ results in larger magnitude coherences, is entirely due to the increased total population, and this is reflected in the steady-state pseudo-entropy. 

As with the previous configuration, we find that $\xi_{12}$ only attains positive values in the transient regime. The other quantum statistical correlations are represented in the table below.
\begin{center}
 \begin{tabular}{||c || c  c || c  c ||} 
 \hline
  & $\chi=$ & $10^{-3}$ & $\chi=$ &  $10^{-2}$  \\ [0.5ex] 
 \hline\hline
 $V(\hat{X}_{1})$ & 0.88, & 13$^{o}$ & 0.67, & 160$^{o}$ \\ 
 \hline
 $V(\hat{X}_{2})$ & 0.74, & 109$^{o}$ & 0.72, & 151$^{o}$ \\ 
 \hline
 DS & 3.9, & 115$^{o}$ & 2.8, & 155$^{o}$ \\  
 \hline
\end{tabular}
\end{center}
We see that the quadrature squeezing results are similar to those of the first configuration, but that there is a significant violation of the Duan-Simon inequality for the higher nonlinearity. This happens because the mode covariances are larger for these parmaters. The steady-state pseudo-entropy was found as $0.02$ for $\chi=10^{-3}$, and $0.03$ for $\chi=10^{-2}$.The actual steady-state reduced density matrices are found as 
\begin{equation}
{\cal R}_{\chi=10^{-3}} = 
\begin{bmatrix}
0.04 & -0.19 \\
-0.19 & 0.96
 \end{bmatrix},
\label{eq:Rg1k3}
\end{equation}
and
\begin{equation}
{\cal R}_{\chi=10^{-2}} = 
\begin{bmatrix}
 0.57 & -0.49 \\
-0.49 & 0.43
 \end{bmatrix}.
\label{eq:Rg1k2}
\end{equation}
With zero coherences and unchanged populations, the values of the pseudo-entropy would be $0.17$ and $0.68$, respectively. We see that, for both configurations, the intracavity systems are far from their closed system equilibrium values. 
The increased violation of the Duan-Simon inequality for the higher non-linearity, and over the first system that we considered, suggest that this system may be the better one for any experimental measurement of bipartite mode entanglement.  

\section{Conclusions}
\label{sec:conclusions}

In conclusion, we have analysed the quantum dynamics of a pumped and damped Bose-Hubbard dimer in two different configurations. Depending on which well is damped, the population dynamics will be very different. The inclusion of a finite collisional term in the equations of motion changes the average solutions from their non-interacting values. In particular, in the second configuration we analysed, with pumping and damping at the same well, collisions cause a finite steady-state population in the first well by contrast to the zero occupation predicted without collisional interaction.

Going beyond the populations, we have found squeezing in the steady-state atomic quadratures, with the amount of squeezing increasing as the collisional nonlinearity is increased. The only configuration for which we found a reasonable entanglement signal between the two wells was for the higher nonlinearity and pumping and damping at different wells. Our calculations of a reduced single-particle pseudo-entropy show that the systems remain far from the equilibrium state of two isolated wells, with the interwell coherences not dropping markedly below those expected for coherent states. Given recent experimental advances, an experimental realisation of these systems should be possible. As a final remark, we note that the truncated Wigner method that we have used easily allows for extension to a greater number of wells. 

\section*{Acknowledgments}

This research was supported by the Australian Research Council under the Future Fellowships Program (Grant ID: FT100100515), and the Conselho Nacional de Desenvolvimento Cient\'{i}fico e Tecnol\'ogico (CNPq).

\end{document}